\pdfoutput=1
\documentclass[11pt]{article}

\usepackage[preprint]{acl}

\usepackage{times}
\usepackage{latexsym}

\usepackage[T1]{fontenc}
\usepackage[utf8]{inputenc}

\usepackage{microtype}

\usepackage{inconsolata}

\usepackage{graphicx}

\usepackage{amsmath} 
\usepackage{amssymb,enumitem} 
\usepackage{multirow}
\usepackage{caption}
\usepackage{wrapfig}
\usepackage{nicefrac}
\usepackage{cleveref}
\usepackage{booktabs}
\usepackage{float}
\usepackage{afterpage}

\title{Towards Federated Low-Rank Adaptation of\\ Language Models with Rank Heterogeneity}

\author{
  Yuji Byun \and Jaeho Lee\\
  Pohang University of Science and Technology (POSTECH) \\
  \texttt{\{yujibyun,jaeho.lee\}@postech.ac.kr} 
}

\begin{document}
\maketitle

\begin{abstract}
%Low-rank adaptation (LoRA) offers an efficient alternative to full-weight adaptation in federated fine-tuning of language models, significantly reducing computational costs. By adjusting ranks for each client, federated LoRA enables flexible resource allocation, and heterogeneous ranks may outperform uniform ones by better matching client resources and data. However, we observe instability due to the conventional zero-padding aggregation strategy, which dilutes information from high-rank clients during model aggregation. To address this issue, we propose a replication-based padding strategy that better retains valuable information from clients with high-quality data. Empirically, this approach accelerates convergence and enhances the global model's performance.

Low-rank adaptation (LoRA) offers an efficient alternative to full-weight adaptation in federated fine-tuning of language models, significantly reducing computational costs. By adjusting ranks for each client, federated LoRA enables flexible resource allocation. However, we observe that heterogeneous ranks among clients lead to unstable performance. Our analysis attributes this instability to the conventional zero-padding aggregation strategy, which dilutes information from high-rank clients during model aggregation. To address this issue, we propose a replication-based padding strategy that better retains valuable information from clients with high-quality data. Empirically, this approach accelerates convergence and enhances the global model's predictive performance.

%Low-rank adaptation (LoRA) is an efficient alternative of full weight tuning for the federated fine-tuning of language models, with significantly lower computational cost. In principle, federated LoRA can allocate different amount of computational budget to each client by tuning ranks for each client. We find, however, that the empirical performance of LoRA is highly unstable with such rank heterogeneity. Our analysis reveals that this instability may be due to the zero-padding-based aggregation strategy adopted in conventional federated LoRA frameworks, which causes the information from high rank clients to get diluted during the aggregation process. To address this issue, we propose a new replication-based padding strategy, which allows us to better leverage the information from clients with high-quality datasets. This method ensures that valuable information from high rank clients is retained during the aggregation process, accelerating the convergence speed and enhancing the overall prediction quality of the global model.

\end{abstract}

\section{Introduction}

Modern language models have shown unprecedentedly strong performance on many tasks \citep{achiam23}, but they also have unprecedentedly many parameters. Their gigantic sizes become especially problematic in \textit{federated fine-tuning} of language models, where the cost to compute and communicate local gradients grows proportionally to the number of parameters \citep{yao2024federated}.

To address this, recent works adopt low-rank adaptation (LoRA; \citet{hu2021lora}) for federated fine-tuning of language models. Instead of tuning all weights, LoRA freezes original weights and trains only the update parametrized as a product of two low-rank matrices. This reduces the number of parameters, thus reducing the computation and communication needed \citep{babakniya2023slora}.

A key promise of federated LoRA is its potential to improve the resource-accuracy tradeoff by adjusting client-wise ranks \citep{cho2023heterogeneous}. Such rank-heterogeneity provides not only a handy way to tune client-wise computation and communication budgets, but also a mean to bias the global update toward certain clients that are considered giving higher-quality gradient estimates.

%Furthermore, by adjusting rank of the local updates, one can easily tune the amount of compute and communication required for each client \citep{cho2023heterogeneous}. Thus, in principle, such rank-heterogeneity can be a useful tool in handling the heterogeneity in data and compute among clients, which is common in real-world scenarios.

In this work, we identify a critical shortcoming of existing rank-heterogeneous federated LoRA methods for language models. Whenever the \textit{quality} of clients varies significantly, conventional rank-heterogeneous LoRA struggles to converge faster than na\"{i}ve rank-homogeneous LoRA. Our analysis suggests that such underperformance might be due to suboptimal \textit{aggregation} strategy; to aggregate LoRA updates with disparate ranks, typical works adopt \textit{zero-padding} strategy, i.e., matching the dimensionality by appending all-zero rows and columns to the low-rank-decomposed parameter updates  \citep{cho2023heterogeneous}. This strategy may not be optimal whenever there exists some clients which provides much higher-quality information, as the information from such clients can be made less relevant by being averaged with padded zeros.

To tackle this problem, we develop a simple yet effective fix, called \textit{replication} strategy. To avoid having highly relevant information from being diluted, we pad lower rank updates with rows and columns replicated from high-priority clients, instead of zeros. Empirically, the proposed method achieves faster convergence to the higher accuracy than existing rank-homogeneous and heterogeneous paradigms. In short, our contributions are:

\begin{itemize}[leftmargin=*,topsep=0pt,parsep=0pt]
   \item We identify the shortcomings of existing rank-heterogeneous federated LoRA frameworks for language models, \textit{i.e.}, unexpected slow convergence under high client quality disparity.
   \item We diagnose the problems in zero-padding-based aggregation, \textit{i.e.,} failing to preserve information from high-quality clients.
   \item We propose a new replication-based aggregation strategy designed to preserve the important information in high-priority clients better, and empirically demonstrate that the proposed method outperforms baseline methods.
\end{itemize}

\section{Background}
\label{background}

\paragraph{LoRA.} LoRA is a parameter-efficient fine-tuning (PEFT) method that keeps the pretrained weights fixed and only trains newly added parameters \citep{hu2021lora}. More concretely, consider fine-tuning a pretrained weight matrix $W_{\mathrm{pre}} \in \mathbb{R}^{m \times n}$. LoRA reparametrizes the updated weight matrix $W_{\mathrm{ft}} \in \mathbb{R}^{m \times n}$ as a sum of the original weight matrix and a product of two low-rank matrices:
\begin{align}
W_{\mathrm{ft}} =  W_{\mathrm{pre}} + BA, \: A \in \mathbb{R}^{r \times n},\:B \in \mathbb{R}^{m \times r}
\end{align}
where $r$ is the rank of the parameter update. As we keep $W_{\mathrm{pre}}$ frozen, only $A$ and $B$ are trainable parameters. Thus, the number of (active) parameters becomes $(m+n)r$, which can be smaller than the number of parameters for the original matrix $mn$ whenever $r$ is sufficiently small. For fine-tuning language models, e.g., LLaMA \citep{touvron2023llama}, it is typical to use $r = 16$ for the matrices of size $m = n = 4096$. In this case, the number of parameter reduces to the $\nicefrac{1}{128} \approx 0.78\%$
 of the original matrix, leading to a proportional decrease in the communication cost for federated fine-tuning.

\paragraph{Federated LoRA.} In federated LoRA with $k$ clients, the server receives $k$ different LoRA updates from the clients. That is, the server receives
\begin{align}
\Delta W_i = B_i A_i,\quad &A_i \in \mathbb{R}^{{r_i} \times n}, B_i \in \mathbb{R}^{m \times {r_i}}
\end{align}
In rank-homogeneous LoRA (\textit{i.e.}, $r_i = r$), a basic way to aggregate the updates from the clients may aggregated by taking an average for both $A$ and $B$ \citep{mcmahan2017communication}. Concretely, one performs
\begin{align}
\bar{A} = \frac{1}{k}\sum_{i=1}^k A_i,\quad \bar{B} = \frac{1}{k}\sum_{i=1}^k B_i\label{eq:loraavg}
\end{align}
The aggregated LoRA weights are then distributed to each client, which is updated further locally until the next communication round.

\paragraph{Zero-padding.} With heterogeneous rank, \textit{i.e.}, whenever $r_i \ne r_j$ does not hold in general, a conventional strategy is to pad the missing dimensions with zero \citep{cho2023heterogeneous}. Concretely, one can consider the zero-padded weight matrices
\begin{align}
\tilde{A}_i^\top &= \big[A_i^\top|\mathbf{0}|\mathbf{0}|\cdots|\mathbf{0}\big] \in \mathbb{R}^{n \times r_{\max}}\nonumber\\
\tilde{B}_i &= \big[B_i|\mathbf{0}|\mathbf{0}|\cdots|\mathbf{0}\big] \in \mathbb{R}^{m \times r_{\max}}\label{eq:zeropadding}
\end{align}
where $r_{\max}$ denotes the maximum rank among all clients. This operation preserves the matrix product $\tilde{B}\tilde{A} = BA$, and thus can be deemed `harmless.' After matching the dimensionality, one can proceed to aggregate the weight updates as in typical rank-homegeneous federated LoRA (\cref{eq:loraavg}).

\section{Shortcomings of the zero-padding}
Our first observation is that the rank-heterogeneous federated LoRA with zero-padding tends to perform worse than rank-homogeneous LoRA, whenever the dataset quality varies significantly over the clients (will be shown later in \Cref{sec:results}, \Cref{fig:distil_albert}). Here, we have varied the dataset quality of each clients by drawing local data from Dirichlet distribution, as in \citet{lin2021fednlp}. Here, the client with larger and more balanced datasets are considered of higher quality, as they achieve higher local accuracy during the early training. We have assigned higher ranks to the higher-quality clients.

\paragraph{Why can zero-paddings hurt?} We hypothesize that such unexpected underperformance of zero-padding is due to the fact that padded zeros tend to dilute useful information captured by high-quality clients. To see this, consider averaging $k$ weight matrices $A_1,\tilde{A}_2,\ldots,\tilde{A}_k$ where $A_1$ is of rank $r_1$ and $\tilde{A}_i$ are of rank $r_2 < r_1$, which is zero-padded with $r_1 - r_2$ all-zero rows. By averaging, the top $r_2$ rows may retain the same relative scale as the original weight. However, the remaining $r_1-r_2$ rows may have the relative scale of $\nicefrac{1}{k}$, having their impact on the overall model much diminished as the number of clients grow.

Indeed, our empirical analysis supports this hypothesis; \Cref{tab:client_a_comparison} compares the accuracy achieved by high-rank clients before and after aggregating the information from low-rank clients. We observe that the accuracy degrades severely after aggregation, suggesting that useful information of the high-rank clients has been lost during aggregation (see \Cref{synthetic_case} for more detailed setup).

\begin{table}[t]
    \setlength{\belowcaptionskip}{-1em}
    \setlength{\aboverulesep}{0.5pt} 
    \setlength{\belowrulesep}{0.5pt} 
    \centering
    \footnotesize
    \begin{tabular}{c@{\hskip 0.05in}c@{\hskip 0.05in}c@{\hskip 0.05in}c@{\hskip 0.05in}c@{\hskip 0.05in}c@{\hskip 0.05in}c} 
        \toprule
        \multirow{2}{*}{High-rank} & \multicolumn{2}{c}{Round 1} & \multicolumn{2}{c}{Round 2} & \multicolumn{2}{c}{Round 3} \\ 
        \cmidrule{2-7} 
        & Before & After & Before & After & Before & After \\ 
        \midrule
        Zero-padding & 84.34 & 38.95 & 71.58 & 42.92 & 86.58 & 50.53 \\ 
        Replication & 84.34 & 82.11 & 88.82 & 86.16 & 89.47 & 86.05 \\ 
        \bottomrule
    \end{tabular}
    \begin{tabular}{c@{\hskip 0.05in}c@{\hskip 0.05in}c@{\hskip 0.05in}c@{\hskip 0.05in}c@{\hskip 0.05in}c@{\hskip 0.05in}c} 
        \toprule
        \multirow{2}{*}{Low-rank (avg.)} & \multicolumn{2}{c}{Round 1} & \multicolumn{2}{c}{Round 2} & \multicolumn{2}{c}{Round 3} \\ 
        \cmidrule{2-7} 
        & Before & After & Before & After & Before & After \\ 
        \midrule
        Zero-padding & 24.96 & 23.95 & 31.07 & 43.42 & 45.06 & 49.11 \\ 
        Replication & 24.96 & 23.95 & 31.07 & 44.08 & 44.48 & 76.63 \\ 
        \bottomrule
    \end{tabular}
    \vspace{-0.5em} 
    \caption{Comparison of accuracy before and after aggregation, for the high-rank client with a high quality local dataset (top) and the low-rank clients that have low quality local datasets (bottom).}
    \label{tab:client_a_comparison}
\end{table}

\section{Method: Replication strategy}

% \noindent\textbf{Understanding the Hazards of Rank Heterogeneity} 
% To understand where the existing paradigm \citep{cho2023heterogeneous}
% falls short, we have tracked the accuracy of client A and other clients after the first three communication rounds. We report the result for the conventional federated LoRA algorithm in table \ref{tab:aggregation comparison zero}. We observe that the performance of the high rank client sharply declines after receiving the aggregated LoRA parameters (which has rank $20$) from the parameter server. This observation suggests that the information relevant to the client A has been lost during the aggregation procedure.

% If so, why does the existing method lose much client A information, despite the fact that the client A contributes much information through rank-20 matrix, while others contribute only 5? To understand why, we need to take a detailed look at the conventional aggregation method.

\begin{figure}[t]
    \centering
    \includegraphics[width=0.9\linewidth]{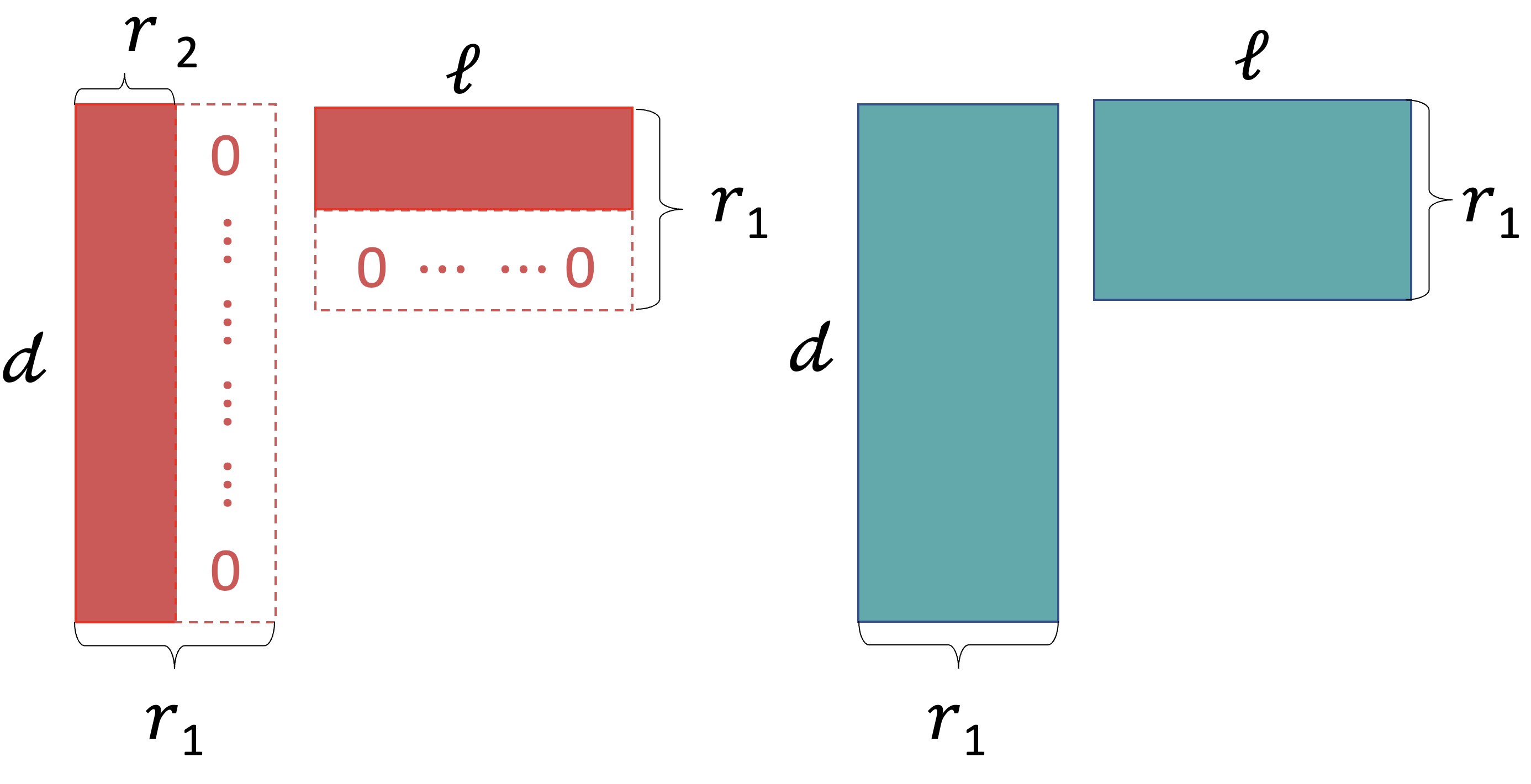} \\
    \includegraphics[width=0.9\linewidth]{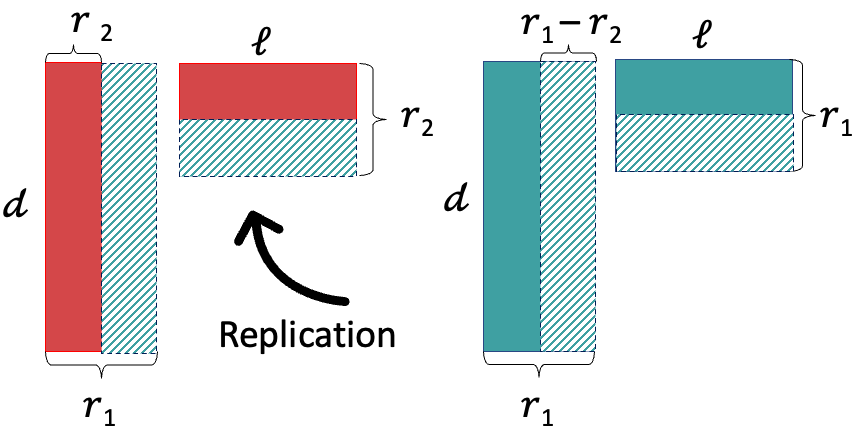}
    \caption{A visual comparison of two strategies for aggregating rank-heterogeneous LoRA updates. Top: Zero-padding. Bottom: Replication (proposed).}
    \vspace{-0.5em} 
    \label{fig:comparison_padding}
\end{figure}

To address this shortcoming, we develop a very simple yet effective method, called \textit{replication} strategy.
%\paragraph{Motivation.} This approach is motivated by the observation that, in many federated learning scenarios, certain clients may possess biased datasets, while others may have access to more diverse datasets; as a result, some clients consistently provide higher-quality updates than others.
%\textcolor{red}{More concretely, consider the case of training a linear model $\hat{y} = Wx$, and a client with a single datum $(\tilde{x},\tilde{y})$. Then the gradient of the squared error loss can be written as $\nabla_{W}(\tilde{y}-W\tilde{x})^2 = (\tilde{y}-W\tilde{x})\tilde{x}^\top$, which is a rank-1 matrix. In other words, a rank-one is sufficient to preserve all gradient information. By a similar reasoning, we can see that the gradients from the samples that lie in a limited number of directions (say, k) can be expressed with at most rank-k updates; the updates for samples in the same direction share the same $\tilde{x}^\top$. In this sense, the data diversity is deeply connected to how large rank is necessary to faithfully preserve the gradient signals. One can extend this example to multi-layer networks, by considering layerwise updates (which are multiplied with succeeding-layer gradients during the backprop). } 
%\textcolor{red}{So, our purpose is by leveraging such differences in data quality across clients, Our strategy aims to preserve the contributions of high-quality clients. To achieve this,} 
Instead of padding all-zero vectors, we replicate the rows and columns from the high-rank clients and append them to low-rank clients (\Cref{fig:comparison_padding}).

Concretely, we first consider a simple case where we have one high-rank client and one low-rank client; let $\Delta W_1 = B_1A_1$ be the high-rank parameter updates from the first client with some rank $r_1$, and let $\Delta W_2 = B_2A_2$ be the low rank parameter update from the second client with rank $r_2 < r_1$. Then, the `row/column-replicated' version of the low rank matrix is given by
\begin{gather}
\tilde{A}_2^\top = \big[ A_2^\top \big|\mathbf{a}^\top_{1,r_2+1} \big| \cdots \big| \mathbf{a}^\top_{1,r_1} \big],\nonumber\\
\tilde{B}_2 = \big[ B_2 \big|\mathbf{b}_{1,r_2+1} \big| \cdots \big| \mathbf{b}_{1,r_1} \big],\label{eq:replication}
\end{gather}
where $\mathbf{a}_{1,i}$ and $\mathbf{b}_{1,i}$ denotes the $i$th row and column vectors of $A_1$ and $B_1$, respectively. Then, we can proceed to aggregating the matrices, as in \cref{eq:loraavg}. Note that the operations can be done rapidly, thus incurring negligible latency to the overall pipeline.

Whenever there are multiple high rank clients, we handle this in three steps: (1) Aggregate high-rank clients (2) Replicate the entries of the aggregated high-rank clients (3) Take a weighted average of the padded low-rank and the aggregated high-rank LoRA updates; here, we set the relative weight of the aggregated high-rank LoRA updates to be proportional to the number of high-rank clients.

We emphasize that the overall communication cost remains unchanged. Since the replication process is performed exclusively on the server, we can enjoy the advantages of our method without any additional communication overhead.

\paragraph{Mechanism for allocating high-rank.} Instead of manually inspecting local datasets to see which client has a high quality dataset (and thus high rank should be allocated), we adopt a simple loss-based criterion to assign high rank. First, we allocate low rank to all clients. After the first local update phase, the server select top-$k$ clients with the highest validation accuracy, and allocate a high rank.

\section{Experimental setup}\label{sec:experiment}
%\subsection{Experimental Setup}

\paragraph{Datasets.} We focus on the text classification, using AG's News \citep{zhang2015character} and DBpedia \citep{auer2007dbpedia} datasets; we preprocess the DBpedia dataset as in \citet{zhang2015character}. We use 10\% of the test set for validation, and the rest for testing.

\begin{figure*}[t]
    \centering    
    \includegraphics[width=0.24\linewidth]{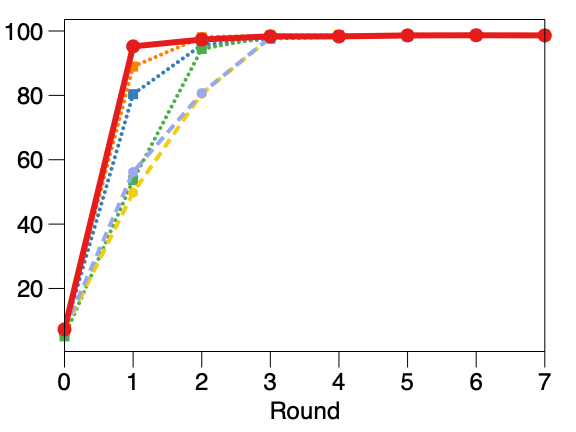} 
    \includegraphics[width=0.24\linewidth]{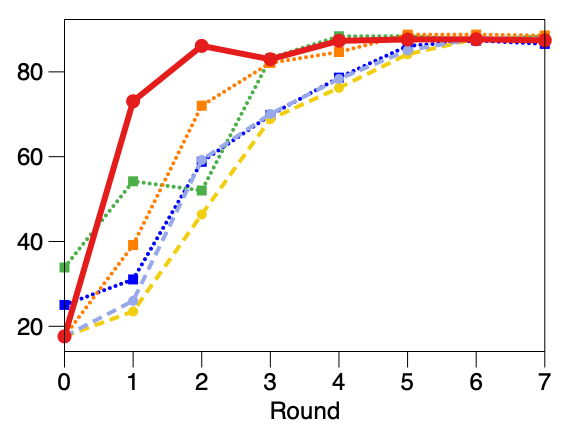}
    \includegraphics[width=0.24\linewidth]{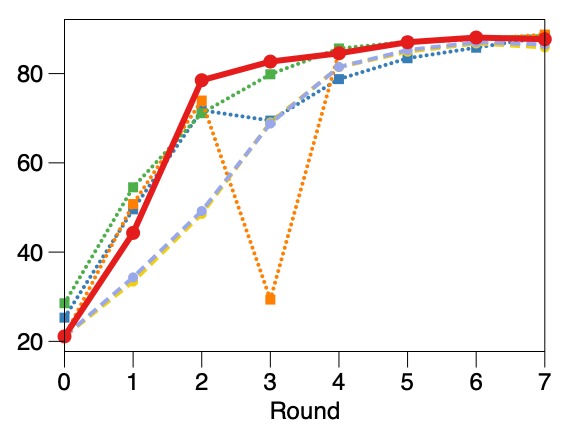} 
    \includegraphics[width=0.24\linewidth]{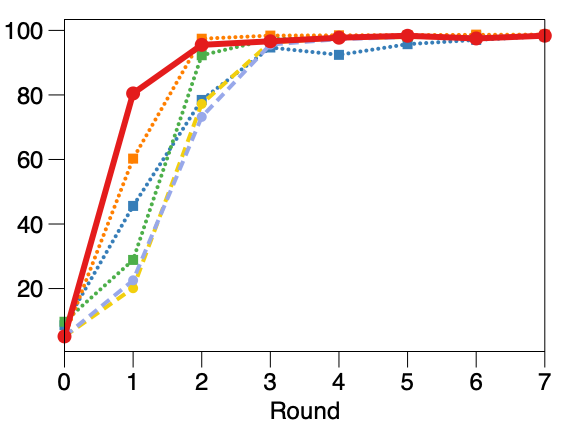}
    \includegraphics[width=1\linewidth]{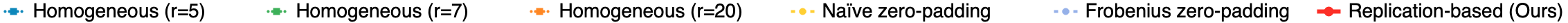} 
    
    \caption{Test accuracy of DistilBERT (left two panels) and ALBERT (right two panels) on the AG's News (first and third) and DBPedia (second and fourth) datasets.}
    \vspace{-0.5em} 
    \label{fig:distil_albert}
\end{figure*}

\paragraph{Models.} We experiment on lightweight BERT-style language models, which are appropriate to be deployed on edge clients: DistilBERT \citep{sanh2019distilbert}, and ALBERT \citep{lan2019albert}. For classification, we add an initialized-and-frozen classification layer to these models, as in \citet{sun2024improving}. We apply LoRA only on self-attention layers, following \citet{hu2021lora}.

\paragraph{Clients.} We employ total $100$ clients, and the training dataset is partitioned over these clients without overlap. We model two types of clients: (1) \textit{High-quality} (HQ) clients have balanced local data, i.e., have similar number of samples for each class. (2) \textit{Low-quality} (LQ) clients have datasets with more class imbalance, i.e., have very few samples from certain classes. We randomly select 10\% of all clients to be HQ, and the remaining 90\% to be LQ. To implement the clients, we follow prior studies \citep{lin2021fednlp,babakniya2023slora} to apply Dirichlet distribution for generating non-\textit{i.i.d.} datasets; we use the hyperparameter $\alpha = \{5.0,1.0\}$ for HQ and LQ, respectively. The average number of samples for both HQ and LQ have been set to be equal. At the initial round, we apply $r=5$ to all clients. After the initial round, we assign $r=20$ to the top 10\% clients that achieve highest validation accuracy.

%\textbf{Datasets and Model.} We conduct a text classification task on AG's News and DbPedia Datasets\cite{zhang2015character}, using DistilBERT \cite{sanh2019distilbert} and ALBERT \cite{lan2019albert} model. And similar to many prior studies\cite{lin2021fednlp},\cite{babakniya2023slora} we apply Dirichlet distribution to account for the non-iid data distribution. The distribution is parameterized by an alpha value, where a lower values indicates higher data heterogeneity, while higher alpha values denote a more uniform distribution. To assume a scenario where specific data possesses relatively uniform distribution, we assigned a 5 alpha value to 10\% of the clients and a 1 alpha value to the remaining clients. Furthermore, we ensured that the average data count per group, differentiated by alpha values, remained consistent.

% \noindent\textbf{Rank assignment.} At the initial round, we apply $r=5$ to all clients. After this round, we assign $r=20$ to the top 10\% clients that achieve highest validation accuracy. 

%In the previous study on Heterogenous LoRA\cite{cho2023heterogeneous}, ranks were distributed using a power-law distribution within the range of \(r_{\text{min}}\) and \(r_{\text{max}}\), but we assigned ranks specifically to two values, \(r_{\text{min}}\) and \(r_{\text{max}}\), to better observe the impact of high ranks.
%In the previous study, training was conducted for up to 200 rounds to observe overfitting.
%However, we determined that this was unnecessary and implemented early stopping, presenting results after 8 rounds, including the initial round.

\paragraph{Training.} Following \citet{mcmahan2017communication}, we conduct one local epoch training per global round. We randomly select 10\% of clients to participate in each global round, ensuring the proportion of high-rank clients remains consistent with the overall distribution. We use Adam with the learning rate 5e-4, without any learning rate scheduling.

\paragraph{Baselines.} We compare the performance of the proposed replication strategy with three baselines. (1) \textit{Homogeneous}: All clients have a same rank, and thus there is no need for an additional aggregation strategy. We evaluate $r \in \{5,7,20\}$, where $r=7$ has a similar total communication cost with the rank-heterogeneous LoRA; see \Cref{tab1} for an explicit communication cost comparison. (2) \textit{Na\"{i}ve zero-padding} \citep{cho2023heterogeneous}: Pads zeros to low-rank updates, as described in \cref{eq:zeropadding}. (3) \textit{Frobenius zero-padding} \citep{cho2023heterogeneous}: Same as na\"{i}ve padding, but applies a weighted sum instead of averaging, with weight proportional to the Frobenius norm of the product matrix $\|\Delta W_i\|_F$.

\section{Results}\label{sec:results}

The experimental results are given in \Cref{fig:distil_albert}.
The leftmost data point denotes the accuracy at initialization (thus can be ignored when comparing baselines), and the subsequent data points denote the test accuracies after each communication round.

\paragraph{DistilBERT.} (Left two) We first observe that the proposed replication strategy (red) achieves the fastest convergence over all methods in both cases. In particular, the strategy closely achieves the peak test accuracy in two communication rounds. In terms of the final accuracy, the proposed strategy is also among one of the best, together with the communication-heavy option (homogeneous rank 20; orange) which only slightly outperforms on AG's News. Zero-padding strategies (dotted lines with circles) converge slower than rank-homogeneous options, with Frobenius padding converging slightly faster than na\"{i}ve. Among rank-homogeneous models, the one with a higher rank tends to converge faster to a higher final accuracy.

\paragraph{ALBERT.} (Right two) Similarly, our method achieves a the fastest convergence to the high accuracy, only slightly worse than the communication-heavy case (homogeneous rank 20). In AG's News, the homogeneous LoRA tend to perform slightly better than the replication-based padding after the very first round; this is because the quality of the high rank client selected in the step by our method happened to be worse than other high rank clients. However, our method quickly starts to outperform the baselines in the subsequent rounds; this suggests that our method performs robust w.r.t. the suboptimalities in the high rank client selection.

\begin{table}[t]
    \centering
    {\footnotesize
    \begin{tabular}{@{\hskip 0.05in}c@{\hskip 0.05in}c@{\hskip 0.05in}c@{\hskip 0.05in}c} 
        \toprule
        & \textbf{\textit{LoRA (r=20)}} & \textbf{\textit{LoRA (r=7)}} & \textbf{\textit{Ours}} \\
        \midrule
        number of parameters & 552,960 & 193,536 & 179,715 \\
        communication cost & 2.11MB & 0.74MB & 0.69MB \\
        fraction of total model & 0.83\% & 0.30\% & 0.27\% \\
        \bottomrule
    \end{tabular}
    }
    \vspace{-0.5em} 
    \caption{Communication cost comparison on DistilBERT. We compare the communication cost used per client (in average) for transmitting LoRA updates.}
    \vspace{-0.5em} 
    \label{tab1}
\end{table}

% \begin{table}[t]
%     \centering
%     {\footnotesize
%     \begin{tabular}{lcccc} 
%         \hline
%         \textbf{Method} & \textbf{\textit{LoRA (r=20)}} & \textbf{\textit{LoRA (r=7)}} & \textbf{\textit{Ours}} & \textbf{\textit{LoRA (r=5)}} \\
%         \hline
%         num of parameters & 552,960 & 193,536 & 179,715 & 138,240 \\
%         communication cost & 2.11MB & 0.74MB & 0.69MB & 0.53MB \\
%         percent of total model & 0.83\% & 0.30\% & 0.27\% & 0.21\% \\
%         \hline
%     \end{tabular}
%     }
%     \caption{Comparison of Communication Costs by Rank with DistilBERT}
%     \label{tab1}
%     \vspace{-1.5em}
% \end{table}

\section{Conclusion}\label{sec:conclusion}

We have identified and analyzed the drawbacks of the zero-padding method during the aggregation process when using heterogeneous LoRA in federated fine-tuning of language models and proposed a replication-based padding method to address these issues. We have experimentally demonstrated that this method achieves faster convergence with lower resource usage compared to homogeneous LoRA with high ranks. This suggests that assigning higher ranks to only a limited set of clients—while leaving others with lower ranks—can better align with client resources and data, optimizing overall performance. Additionally, this study focuses on a single high rank and a single low rank, allowing for exploration of multiple ranks to better manage resource and data heterogeneity. We believe that our research opens up new challenges and opportunities in federated fine-tuning, and we are confident that this study will contribute to more efficient federated learning.

%\noindent\textbf{Limitation.} Our approach significantly leverages the influence of high-performing clients to enhance the performance of the global model. As mentioned in the experimental results, there is a risk of overall performance degradation if a high rank client with relatively poor performance is selected during the client selection process. However, in our experiments, we used a simple random selection method and recent studies have explored selecting clients while considering data heterogeneity. We believe that applying more sophisticated client selection methods can mitigate this limitation. 

%\noindent\textbf{Future work.} Currently, we consider a scenario with a single high rank and a single low rank, i.e., two ranks. However, it is also possible to explore methods that apply multiple ranks, taking data heterogeneity into account. Additionally, just as we replaced zero-padding with replication-based padding, we could consider alternative methods to replace truncation. Finally, while we are currently padding and aggregating by aligning the same rows and columns of low rank and high rank clients, we can also explore other efficient methods to match rows and columns between low rank and high rank clients.

\section*{Acknowledgments}
This work has been supported by the National Research Foundation of Korea (NRF) grant funded by the Korea government (MSIT) (No. RS-2023-00213710, No. RS-2024-00453301).

\section*{Limitations}\label{sec:limit}
Our approach is based on the assumption that at least one client possesses high-quality data in the federated learning setting. In cases where all clients have data of similarly high quality, the performance gains of our method may be limited. In addition, we have only explored a binary categorization of clients (high-quality, and low-quality), while in practice the client quality can be quite diverse.  %Also, the quality of client data is determined solely during the rank assignment process in the initial rounds. If a low-quality (LQ) client gets \textit{lucky} and perform well in these initial rounds, it could negatively influence the subsequent federated learning process.

\bibliography{references}

\appendix
\label{sec:appendix}

\begin{figure*}[!ht]
    \vspace{-1em} 
    \centering    
    \includegraphics[width=0.26\linewidth]{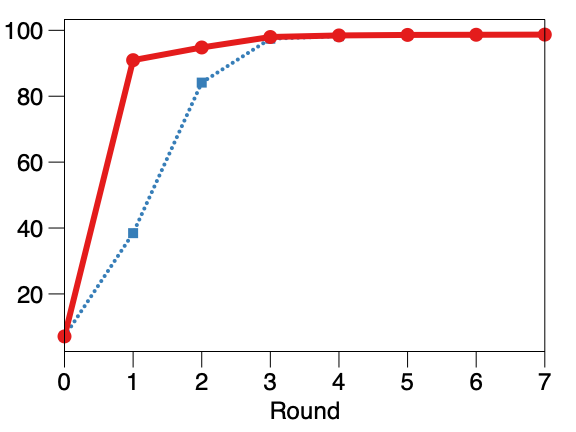} 
    \hspace{0.5cm}
    \includegraphics[width=0.26\linewidth]{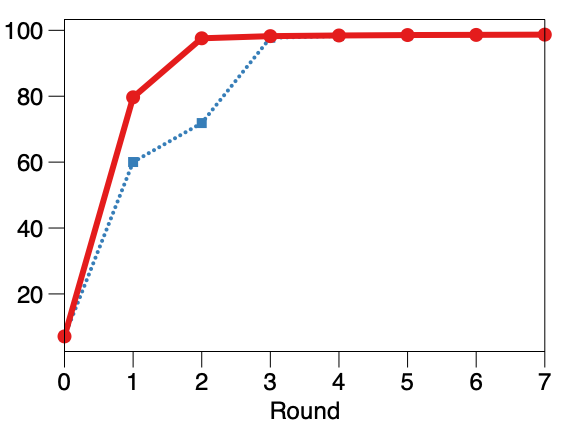}
    \hspace{0.5cm}
    \includegraphics[width=0.26\linewidth]{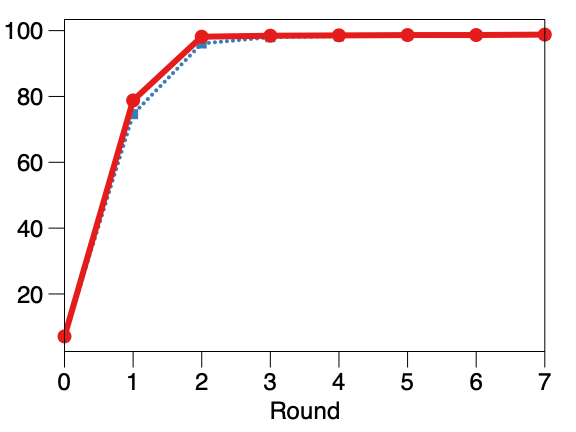} 
    \includegraphics[width=0.35\linewidth]{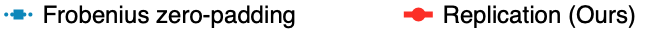} 
    
    \caption{Test accuracy based on the proportion of high-rank clients, with the results shown for 10\%, 20\%, 50\% of high-rank clients from left to right.}
    \vspace{-0.5em} 
    \label{fig:propotion}
\end{figure*}

\section{Related work}
Data heterogeneity, or the discrepancy among the client-wise data distribution, has been studied extensively in federated learning. Such heterogeneity is very common in real world scenarios, and can severely degrade the model performance \citep{zhu2021federated}. Many works have focused on resolving this issue, proposing various solutions including that involve data sharing \citep{zhao2018federated} or better calibration of batch normalization \citep{li2021fedbn}.

The dataset heterogeneity has also been discussed in the context of parameter-efficient federated learning as well. For instance, \citet{kim2023client} studies how the negative impacts of dataset heterogeneity can be mitigated the federated learning of adapters \citep{houlsby2019parameter}.
% SLoRA \citep{babakniya2023slora} applies LoRA to the federated learning scenario, and proposes a refined initialization scheme for resolving the dataset heterogeneity. 
Most closely related to our work, \citet{cho2023heterogeneous} considers assigning different rank for the clients, as a mean of addressing inter-client heterogeneity.

In contrast to these works, our work primarily focuses on the scenario where the \textit{relative importance} of each client can be dramatically different. Clients with similar data distribution can have very different importances whenever the amount of data significantly differs, and vice versa when both clients have similar degree of imbalance with different majority classes. When some clients are notably of better quality than others, we demonstrate that the algorithm of \citet{cho2023heterogeneous} may not be effective; our work proposes a way to fix this problem.

\section{Experimental setup for Table \ref{tab:client_a_comparison}}
\label{synthetic_case}
To establish a simple experimental setup, We conduct the experiments using DistilBERT and AG's News dataset and considered 15 clients. One client ha
d a perfectly uniform data distribution, while the remaining clients followed a Dirichlet distribution with $\alpha = 0.6$, the average number of data points from these clients has been kept equal to the number of data points of the client with uniform distribution.

\section{Additional experiments}
For additional discussion, We conduct the experiments using the DistilBERT model and the DBPedia dataset. These experiments focus on examining the effects of rank allocation and varying the proportion of high-rank clients.

\begin{figure}[!ht]
    \centering    
    \includegraphics[width=0.6\linewidth]{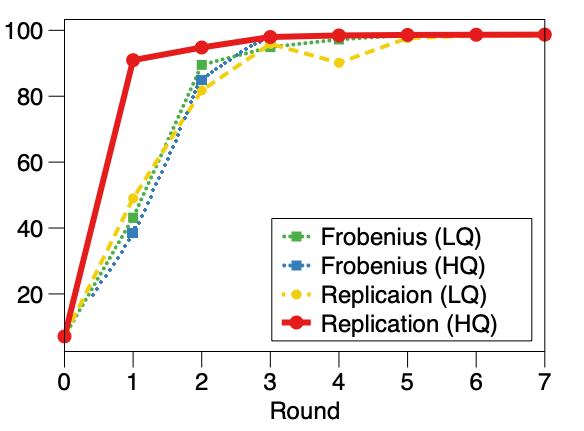} 
    \caption{Comparison of model performance based on rank allocation}
    \label{fig:hqlq}
    \vspace{-0.5cm} 
\end{figure}

\subsection{Rank allocation}
\vspace{0cm} 
To demonstrate the advantages of assigning high ranks to clients with high-quality data, we conduct an experiment where high ranks were assigned to clients with low-quality data, specifically those in the bottom 10\% in the initial round.  As shown in Figure \ref{fig:hqlq}, We observe that assigning high ranks to low-quality clients did not result in better performance than even simple Frobenius zero-padding. This suggests that copying the weights of models trained on imbalanced data offers limited benefits.

\subsection{Proportion of high rank clients}
\vspace{0cm} 

To compare results based on the high-rank client ratio, we conduct experiments with high-rank client ratios set at 10\%, 20\%, and 50\%. The results can be seen in Figure \ref{fig:propotion}. As the high-rank client ratio increases, the performance gap with Frobenius zero-padding diminishes. This trend can be interpreted as the disadvantage of diluting high-rank information being offset by the reduction in replicated weights. However, it is important to note that as the proportion of high-rank clients increases, more resources are required.

\section{Other experimental details}
 All experiments were executed on a single NVIDIA RTX A6000 GPU without distributed training. The graphs within the figure were generated using a single fixed random seed for consistency.

\end{document}